\documentclass[10pt]{article}
  \usepackage{graphicx}
  \DeclareGraphicsExtensions{.eps}
\begin{document}
\title{Point-to-point and Point-to-multipoint CDMA Access Network with Enhanced Security}
\author{
  Alfredo~A.~Ortega,
  V\'{\i}ctor~A.~Bettachini,
  \thanks{A. A. Ortega, V. A. Bettachini, J. I. Alvarez-Hamelin, and D.
F. Grosz are with Instituto Tecnol\'ogico de Buenos Aires, 25 de Mayo
444, C1002ABJ, Buenos Aires, Argentina (e-mail: aortega@alu.itba.edu.ar,
\{vbettachini,ihameli,dgrosz\}@itba.edu.ar).}\\
  Jos\'e~Ignacio~Alvarez-Hamelin
  and Diego~F.~Grosz,%
  \thanks{J. I. Alvarez-Hamelin, and D. F. Grosz are also with CONICET
(Argentine Council of Scientific and Technological Research).}
}
\maketitle
\begin{abstract}
We propose a network implementation with enhanced security at the physical layer by means of time-hopping CDMA, supporting cryptographically secure point-to-point and point-to-multipoint communication.
In particular, we analyze an active star topology optical network implementation capable of supporting $128$ simultaneous users up to $20$ km apart.
The feasibility of the proposed scheme is demonstrated through numerical simulation.
{\bf keywords:} {\em optical fiber communication, access networks,
secure communication, CDMA}
\end{abstract}
\section{Introduction}

Communication methods employed today in access optical
networks are inherently insecure, as they do not provide privacy from a
sufficiently motivated eavesdropper, as signals are broadcasted to all
users, e.g., as in Passive Optical Networks (PONs).
Communication security must be implemented by the end-points on higher-level protocols, and is often neglected.

This paper proposes an access network architecture where security is provided by a time-hopping CDMA scheme at the physical layer.
The proposed scheme provides point-to-point and point-to-multipoint
communication, allowing the setup of Virtual Private Networks (VPNs)
among users.
In this way, each client transmits in frame slots assigned by a
cryptographically secure PRBS.  Since each user transmits one bit per
slot, collisions occur and are
handled by a combination of Bloom
filters~\cite{Bloom70space/timetrade-offs} and
standard error-correction algorithms optimized for Z channels~\cite{Golomb:80}.
Bloom filtering is a technique borrowed from hashing algorithms where a
bit is transmitted in $K$ slots into the same frame; therefore it is
sufficient to receive a single `0' out of $K$ copies in order to
correctly retrieve the original transmitted `0' whereas, in a Z
channel, collisions have no effect on `1s'.

An early approach to this idea was presented in \cite{Tancevski:95},
where a combination of wavelength hopping and direct sequence CDMA
with orthogonal codes was used to provide security in the link, whereas our proposal uses a single wavelength and time hopping CDMA.
More recently, Refs. \cite{Nadarajah2006} and~\cite{Nirmalathas2009} proposed a VPN built with direct sequence CDMA using Walsh orthogonal coding on a single wavelength channel.
Although orthogonal codes make an efficient use of the available
bandwidth, in Ref.~\cite{Shake:05} it is shown that the security they
provide is weak due to the reduced code space.
Finally, Ref. \cite{Wang:10} used the Advanced Encryption Standard (AES)
to encrypt data, and then sent this encoded stream by direct sequence CDMA with
a short code length, that is, security is provided in a higher
layer, i.e., as is standard cryptography~\cite{Shake:05}.
In contrast, our proposed scheme uses a $key$ to feed each PRBS giving
 non-orthogonal coding sequences, thus producing a larger search space
in case an attack is attempted. Moreover we use a non-linear generator,
e.g., a self-shrinking generator~\cite{Meier:94}.

The architecture proposed in this paper consists of a star network
topology using a single wavelength, in contrast to other optical access network designs \cite{journals/jsac/CarenaFFGNPP04}.
The proposed topology and functionality resemble that of a PON (users
can be regarded as Optical Network Units, ONUs) but note that
point-to-point as well as point-to-multipoint communication are
supported. 
To receive an specific channel, the user/ONU needs the corresponding
{\em key}, thus it can communicate as many users as {\em keys} it has.
A privileged user can communicate all other users acting as the Optical Line Termination (OLT) in PONs.

\section{Architecture}
\begin{figure}[!t]
  \centering
    \includegraphics[width=3in]{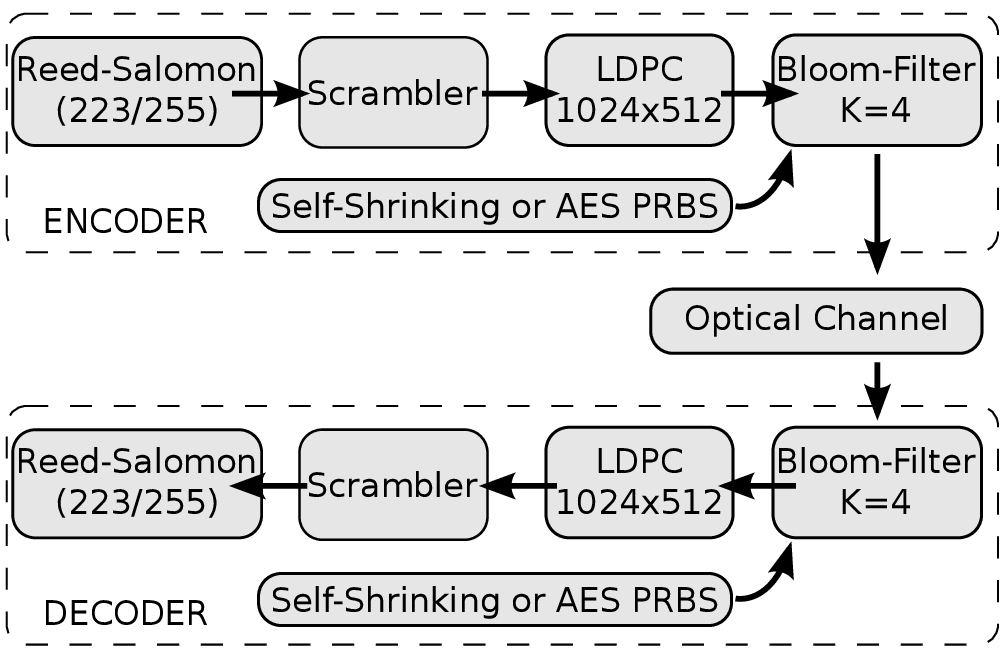}
    \caption{Proposed network design: Access Layer}
    \label{arch:chain}
\end{figure}

The proposed system is composed of an access layer, where CDMA and error correction are implemented, and a physical layer based on an optical network with certain similarities to PONs. 
The access layer is implemented using time-hopping CDMA, where each of the $128$ possible ONUs sends bits in a slot randomly chosen from a frame of $356$ slots; therefore
collisions between different ONUs will occur and error correction must be used to guarantee error-free transmission. 
Notice that the synchronization is performed at the bit slot level only
because transmission of each ONU is random, in contrast to TDMA where
synchronization is also performed at the frame level. 
Moreover, each ONU can send data at any time, in contrast to TDMA where
ONUs usually send data continuously; this feature resembles transport by Ethernet frames.
A certain ONU $X$ can receive messages from an ONU $Y$ if $X$ has the
{\em key} of $Y$, and viceversa. Therefore, if a certain group of ONUs
were to communicate over a VLAN, it is required that everyone in the group
knows each others' {\em keys}.
ONUs' data streams are encoded with the following error correction techniques (Fig. \ref{arch:chain}):
 Reed-Solomon ($223/255$) and LDPC ($1024\times512$ matrix) algorithms (see \cite{Moon:05} and references therein), and Bloom-filters with $K=4$~\cite{Bloom70space/timetrade-offs}.
The choice of these correction algorithms was heavy influenced by
modeling the optical fiber as a Z-channel, having a Shannon limit of $ C_{Z} = \log_2\left(1+(1-p) p^{p/(1-p)}\right),$ where $p$ is the probability of error. 
This capacity limit is larger than that of a symmetric memory-less binary channel \cite{Tallini:02}.

\begin{figure}[!t]
  \centering
    \includegraphics[width=3in]{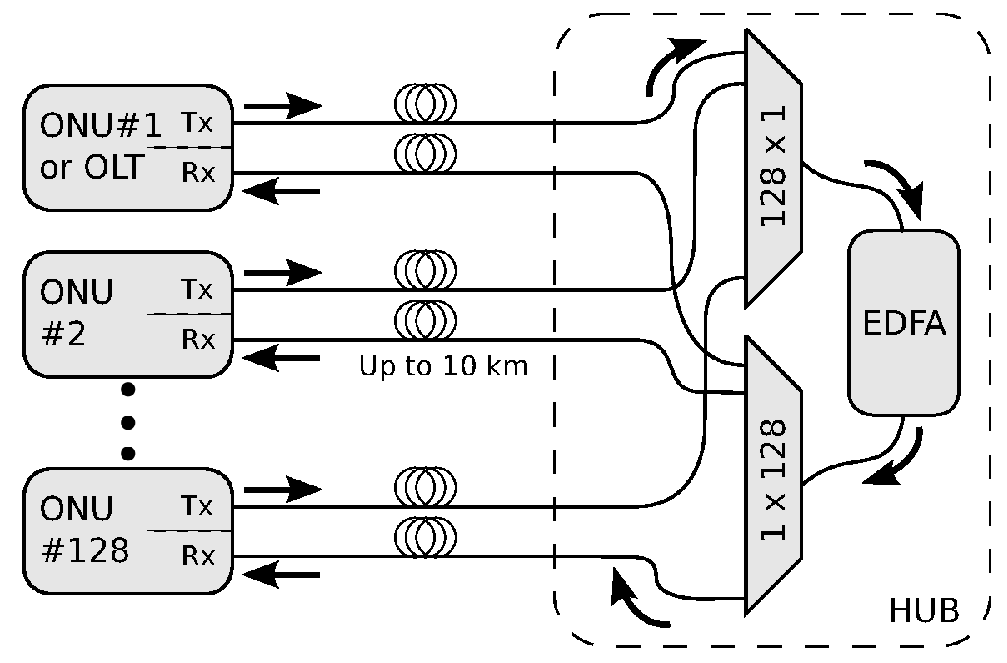}
    \caption{Proposed network design: Optical Layer}
    \label{arch:fig1}
\end{figure}

The proposed physical layer topology is that of a star (see Fig.
\ref{arch:fig1}) where optical splitters redistribute traffic coming
from each ONU to all the rest allowing point-to-multipoint as well as
point-to-point communications between $128$ ONUs.
An Erbium-Doped Fiber Amplifier (EDFA) located in between splitters at
the optical hub increases optical power to overcome network losses.  RZ
modulated optical signals generated at each ONU, of up to $10$~Gbps by a
$2$~dBm $1550$~nm DFB-laser, are transmitted up to $10$~km upstream by a
standard single-mode optical fiber (ITU-T G.652) to the optical hub.

In this hub a $128\times 1$ splitter merges traffic from all ONUs that is then redistributed by a $1\times 128$ splitter channeling back merged traffic to each ONU through a downstream fiber identical and parallel to the upstream fiber.
Splitters' attenuation ($\simeq25$~dB each) contribute, as well as fiber attenuation and insertion losses ($\simeq2$~dB and $\simeq1$~dB per stretch), to high total losses ($\simeq28$~dB at both upstream and downstream paths).
In order to provide signal amplification an EDFA ($\geq27$~dB gain and noise figure $7$~dB) is placed between both splitters.
This EDFA increases merged traffic power at the first splitter output ($\simeq-26$~dBm `1' active Tx) delivering an adequate power level ($1$~dBm, `1' active Tx) at the second splitter input to provide ONU's receiver a power level for proper reception ($-27$~dBm, `1' active Tx) with a high sensitivity ($-28$~dBm) photodetector (PD).
The PD maximum optical power is not a concern as our simulations show that only up to ten `1s' collide at any given bit slot.
Even considering a constant EDFA gain, the PD input optical power would be lower ($-17$~dBm) than that commercial PDs withstand unharmed ($\sim -5$~dBm).
The bit `0' level at PD is given by the addition of the `0' bit transmitted 
by all $128$~ONUs.
The receiver decision threshold should be able to separate between this state and that of a single ONU transmitting a `1' bit.
As the bit `0' transmission power should be very low, imposing
restrictions on the DFB-laser extinction ratio.
The minimal required extinction ratio (`1'$/$`0' peak power ratio) is addressed in the numerical simulations explained next.

\section{Numerical simulations}
We developed a modular simulator platform both for ECC and optical channel stages.
It was released under the GNU license~\cite{sim1}.

The physical optical channel simulation block provides an estimate of
the BER performance of the optical channel. Simulation steps are as
follows: RZ upstream traffic coming from all ONUs is assumed to arrive
at the $128\times1$ splitter with perfect time synchronization, i.e., there is no timing jitter. 
The `0'-bit slots contain a small CW optical intensity given by the Tx extinction ratio. 
Each on-line ONU adds its `0'-bit optical intensity yielding a base power level.
Each `1'-bit adds a super-Gaussian ($m=4$) pulse, duty cycle $1/3$, to the base power level. 
 
Upstream and downstream merged traffic suffers from attenuation due to
splitter, fiber, and
splice losses. The power budget is balanced by an EDFA with $27$~dB constant gain.
Amplified spontaneous emission from the EDFA is modeled by white Gaussian
noise, with intensity proportional to the amplifier noise figure ($7$~dB), and is added
after the EDFA.

The input optical signal at the receiver is filtered (2nd order low-pass Butterworth filter, $25$~GHz bandwidth) and photodetected assuming a standard PD responsivity (see section 4.4.3 of~\cite{Agrawal:xx}).
White Gaussian noise accounting for thermal and shot noise is then added
to the photocurrent, and 
electrical filtering is applied (2nd order low-pass Butterworth filter, $14$~GHz bandwidth).

\begin{figure}[!t]
    \centering
      \includegraphics[angle= 270, width=3.5 in]{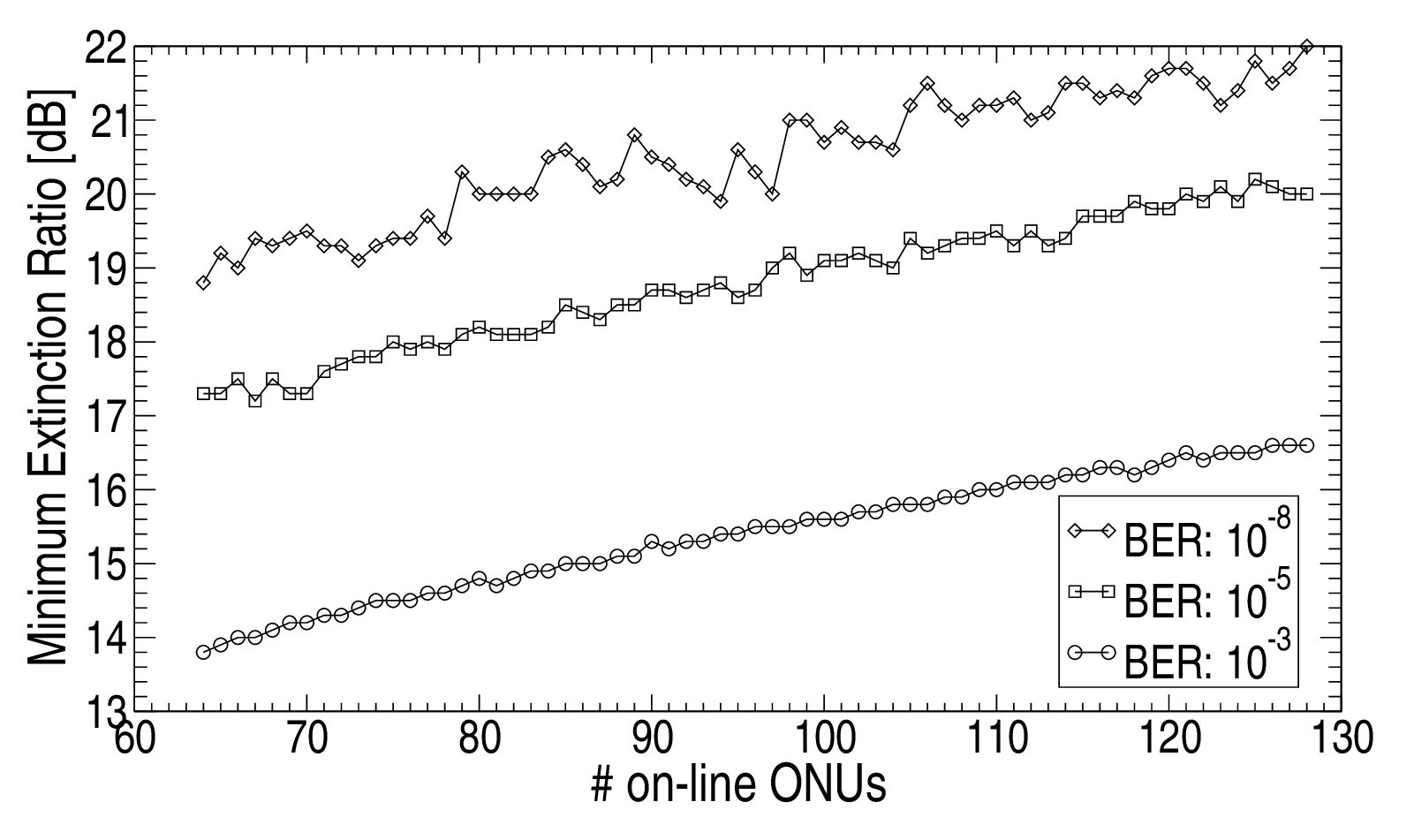}
      \caption{Physical (optical) layer simulation result: Minimal extinction ratio required to assure a given BER.}
      \label{sim:optical}
\end{figure}

Noise fluctuations at power levels near the PD sensitivity limit have an important effect on signal detection. 
Shot noise is of particular concern as it is proportional to the mean photocurrent.
In our network proposal the later is higher than in PONs as bit
`0' optical intensities from all ONUs are added.
The resulting base-level optical intensity is then heavily dependent on the Tx extinction ratio.
Fig.~\ref{sim:optical} shows minimal extinction ratios required to
achieve an arbitrary BER in the physical layer as a function of the
number of on-line ONUs.
In the $128$ ONUs scenario a  BER$<10^{-3}$ can be achieved using
commercially available transmitters with an extinction ratio $\simeq16.6$~dB.
This BER is low enough to allow for logical-channel error-correction routines that guarantee error-free transmission, while still making use of a fair fraction of channel capacity when non-orthogonal codes are used.
\begin{figure}[!t]
    \centering
      \includegraphics[angle=270 ,width=3.5 in]{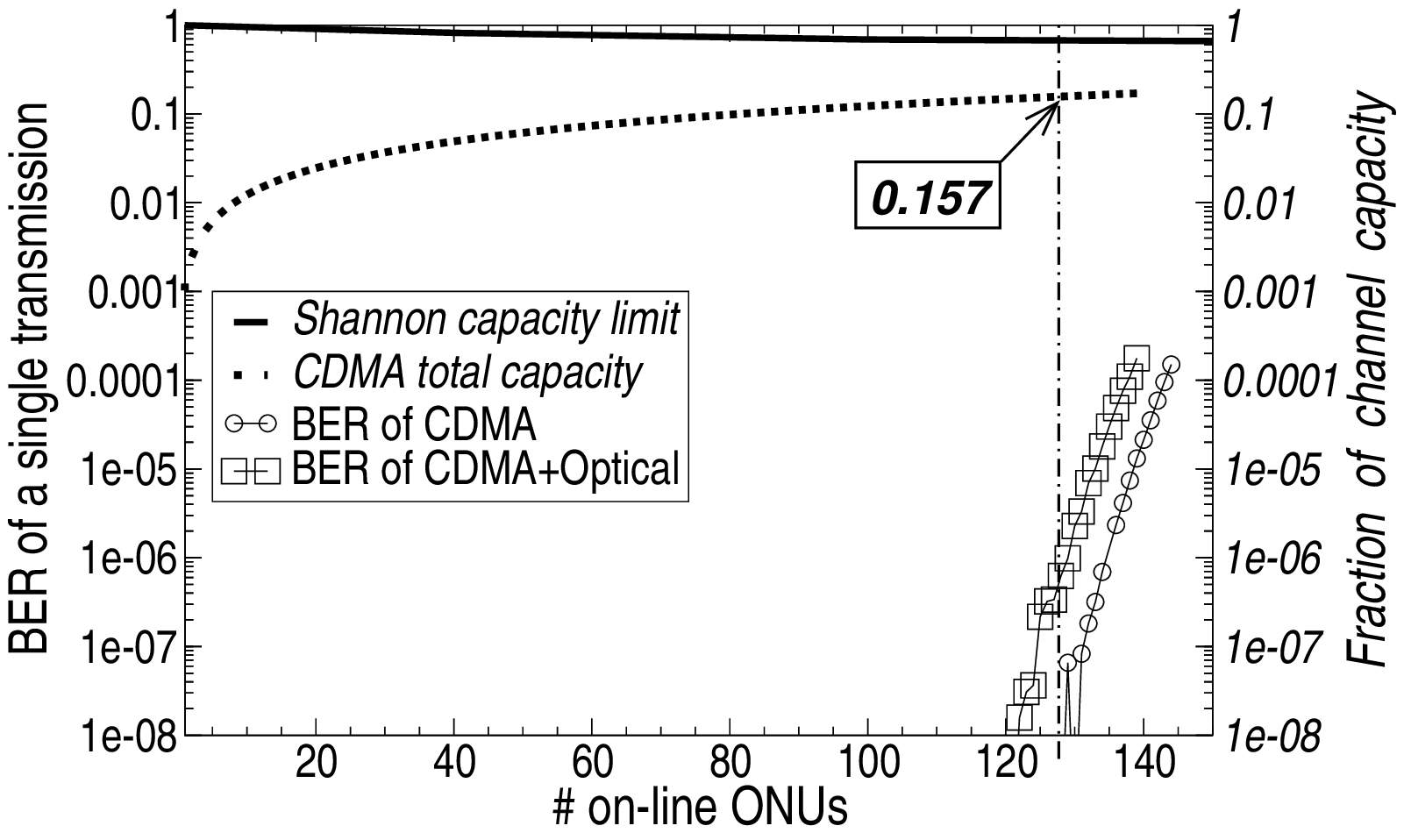}
    \caption{Simulation results: Logical channel}
      \label{sim:access}
\end{figure}
Fig.~\ref{sim:access} shows simulation results for the fraction of the total
capacity and the BER of one channel at the coding level (circles) and 
including physical layer impairments (squares). These results were obtained by
sending one Gigabit of data for each ONU simultaneously.
This figure shows a channel utilization of $15.7\%$ when all of $128$ ONUs
are transmitting simultaneously, with a BER$<10^{-8}$. 
From Fig. \ref{arch:fig1} we observe a penalty of $8$ ONUs when
impairments from the optical layer (mainly extinction ratio and noise from EDFA and PDs) are taken into account.
Considering that the system was designed to support asynchronous communications (e.g., Ethernet), it is not likely that all the ONUs will transmit simultaneously (e.g., Internet links often operate at most at $90\%$ load); and therefore our system has a BER $<10^{-8}$ for each channel when 119 ONUs are transmitting at a same time ($119/128>0.9$).
Observe that the high error rates correspond to a
worst-case scenario when all ONUs are transmitting simultaneously at
full capacity, and also 
there is a low penalty due to physical layer impairments.

\section{Security Performance}
There are four basic goals related to security performance: Confidentiality, integrity, availability and authenticity \cite{Dhillon:07}.
The proposed scheme provides confidentiality and integrity in the downstream channel, leaving availability and authenticity implementations to higher-level protocols. Confidentiality and integrity are achieved by using time-hopping CDMA.
Attacks on this kind of systems are extensively studied in~\cite{Shake:05}. The proposed system is resistant to brute-force attacks since the code space size can be grater than $2^{256}$ (related to the period of the PRBS), the proposed PRBS is non-linear e.g., a self-shrinking generator~\cite{Meier:94} or AES in PRBS mode, and the key of each ONU is never broadcasted to the network. The key distribution must be performed beforehand using any secure method available~\cite{Denning:82}

\section{Conclusions}
We proposed a time-hopping CDMA network architecture capable of
supporting both point-to-point and point-to-multipoint communication of
up to $128$ ONUs with a worst-case rate of $12$~Mbps. Furthermore, the
proposed scheme provides security at the physical layer and VPNs between
ONUs can be set up without additional higher-level protocols.  We also
showed that there is a low penalty due to physical
layer
impairments, such as transmitter extinction ratio and attenuation at
fibers, splitter, and splices.
We believe that our proposal opens the door to the design of a larger
secure
CDMA network, covering longer distances and servicing more end users.

\section*{Acknowledgment}
This work was supported by grant PICT-497/2006 ANPCyT Argentina, and also by Core Security Technologies.

\bibliographystyle{plain}
\bibliography{IEEEabrv,IEEEorte}
\end{document}